\documentclass[aps,prl,floatfix,twocolumn,reprint,amsmath,amssymb,superscriptaddress]{revtex4-1}
\usepackage{amsfonts}
\usepackage{mathrsfs}
\usepackage{amsmath}
\usepackage{color}
\usepackage{natbib}
\usepackage{graphicx}
\usepackage{bm}
\usepackage{amssymb}
\usepackage{xspace}
\usepackage{epstopdf}
\usepackage{dcolumn}
\usepackage{longtable}
\usepackage{multirow}
\usepackage[colorlinks=true, letterpaper=true, pdfstartview=FitV, linkcolor=blue, citecolor=blue, urlcolor=blue]{hyperref}

\makeatletter

\newcommand{\Rmnum}[1]{\expandafter\@slowromancap\romannumeral #1@}
\makeatother

\begin{document}
\title{Type-II nodal loops: theory and material realization}

\author{Si Li}
\affiliation{Beijing Key Laboratory of Nanophotonics and Ultrafine Optoelectronic Systems, School of Physics,
Beijing Institute of Technology, Beijing 100081, China}
\affiliation{Research Laboratory for Quantum Materials, Singapore University of Technology and Design, Singapore 487372, Singapore}

\author{Zhi-Ming Yu}
\affiliation{Research Laboratory for Quantum Materials, Singapore University of Technology and Design, Singapore 487372, Singapore}

\author{Ying Liu}
\affiliation{Research Laboratory for Quantum Materials, Singapore University of Technology and Design, Singapore 487372, Singapore}

\author{Shan Guan}
\affiliation{Beijing Key Laboratory of Nanophotonics and Ultrafine Optoelectronic Systems, School of Physics,
Beijing Institute of Technology, Beijing 100081, China}
\affiliation{Research Laboratory for Quantum Materials, Singapore University of Technology and Design, Singapore 487372, Singapore}

\author{Shan-Shan Wang}
\affiliation{Research Laboratory for Quantum Materials, Singapore University of Technology and Design, Singapore 487372, Singapore}

\author{Xiaoming Zhang}
\affiliation{Research Laboratory for Quantum Materials, Singapore University of Technology and Design, Singapore 487372, Singapore}

\author{Yugui Yao}\email{ygyao@bit.edu.cn}
\affiliation{Beijing Key Laboratory of Nanophotonics and Ultrafine Optoelectronic Systems, School of Physics,
Beijing Institute of Technology, Beijing 100081, China}

\author{Shengyuan A. Yang}\email{shengyuan\_yang@sutd.edu.sg}
\affiliation{Research Laboratory for Quantum Materials, Singapore University of Technology and Design, Singapore 487372, Singapore}

\begin{abstract}
Nodal loop appears when two bands, typically one electron-like and one hole-like, are crossing each other linearly along a one-dimensional manifold in the reciprocal space. Here we propose a new type of nodal loop which emerges from crossing between two bands which are both electron-like (or hole-like) along certain direction. Close to any point on such loop (dubbed as a type-II nodal loop), the linear spectrum is strongly tilted and tipped over along one transverse direction, leading to marked differences in magnetic, optical, and transport responses compared with the conventional (type-I) nodal loops. We show that the compound K$_4$P$_3$ is an example that hosts a pair of type-II nodal loops close to the Fermi level. Each loop traverses the whole Brillouin zone, hence can only be annihilated in pair when symmetry is preserved. The symmetry and topological protections of the loops as well as the associated surface states are discussed.
\end{abstract}
\maketitle

Topological metals and semimetals have become a focus of current physics research~\cite{Burkov2016,Yan2017}. These materials feature nontrivial band-crossings in their low-energy band structures, around which the quasiparticles behave drastically different from the usual Schr\"{o}dinger-type fermions. Depending on its dimensionality, the crossing manifold may take zero-dimensional (nodal point), one-dimensional (nodal loop), or two-dimensional (nodal surface) form~\cite{Yang2016}. There has already been extensive studies on nodal points, especially on so-called Weyl and Dirac semimetal materials~\cite{Wan2011,Murakami2007,Burkov2011,Volovik2003,Young2012,Wang2012b,Zhao2013c,Yang2014a,Weng2015,Huang2015,Liu2014c,Borisenko2014,Lv2015,Xu2015a}. Recently, nodal loops begin to attract considerable interest: several nodal-loop materials have been proposed, with interesting physical consequences revealed~\cite{Weng2015c,Yang2014c,Mullen2015,Yu2015,Kim2015a,Chen2015,Xie2015,Fang2015,Chan2016,Li2016,Bian2016,Schoop2016,Fang2016,HXu,Gan,Yu2017}.

\begin{figure}[b!]
\includegraphics[width=9cm]{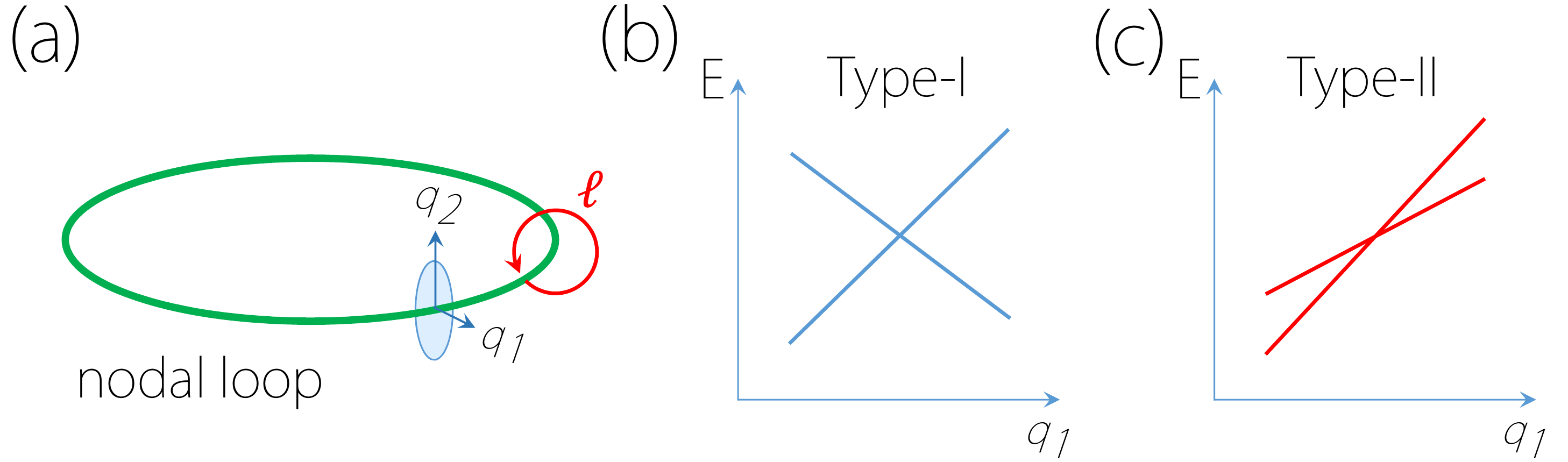}
\caption{(a) Schematic figure of a nodal loop. $q_1$ and $q_2$ label the two transverse directions. (b) and (c) illustrate the type-I and type-II dispersions along the $q_1$ direction.  }
\label{fig1}
\end{figure}

Consider the generic case of a nodal loop formed by the linear crossing between two bands in a three-dimensional system. Close to any point $P$ on the loop, the dispersion is linear along the two transverse directions of the loop, and is at least quadratic along the tangential direction. The low-energy effective model near $P$ can be expressed as (set $\hbar=1$)
\begin{equation}\label{Pmodel}
\mathcal{H}=v_1 q_1 \sigma_x+v_2 q_2 \sigma_y+\bm w\cdot\bm q,
\end{equation}
up to first order in the wave-vector $\bm q$ measured from $P$. Here $q_i$'s ($i=1,2$) are the components of $\bm q$ along two orthogonal transverse directions [see Fig.~\ref{fig1}(a)], $v_i$'s are the Fermi velocities, and $\sigma$'s are Pauli matrices denoting the two-band degree of freedom. The last term with a vector $\bm w$ in (\ref{Pmodel}) represents a tilt of the spectrum, such that the energies of the two eigenstates $|u_\pm (\bm q)\rangle$ are given by
\begin{equation}\label{Epm}
E_{\pm}=\bm w\cdot\bm q\pm \sqrt{v_1^2 q_1^2+v_2^2 q_2^2}.
\end{equation}
In the $q_1$-$q_2$ plane, the tilt is most effective along the $\bm w_\bot$ direction, where $\bm w_\bot=(w_1,w_2,0)$ is the projection of $\bm w$ onto the $q_1$-$q_2$ plane.

When $|\bm w_\bot|$ is small, the spectrum shows a usual band-crossing pattern for conventional nodal loops [see Fig.~\ref{fig1}(b)]: the crossing is of linear-type, and the slopes of two bands have opposite signs for \emph{all} directions in the  $q_1$-$q_2$ plane.  The spectrum is then fully gapped along a small loop $\ell$ encircling the nodal loop [Fig.~\ref{fig1}(a)], such that $\ell$ is characterized by a $\pi$ Berry phase:
$
\oint_\ell \langle u_-|i\nabla_{\bm q}u_-\rangle\cdot d\bm q=\pi
$,
which may be intuitively understood by tracing the winding of the pseudospin $\bm\sigma$ for the lower band when moving around $\ell$.
When the system possesses both time reversal ($\mathcal{T}$) and inversion ($\mathcal{P}$) symmetries, and when the spin-orbit coupling (SOC) can be neglected, the Berry phase along any closed loop must be quantized in unit of $\pi$~\cite{Schnyder2008}, providing a topological protection of the nodal loop from gap-opening. Another commonly encountered protection mechanism comes from mirror reflection symmetry: a nodal loop in a mirror-invariant plane is protected when the two crossing bands have opposite mirror eigenvalues.

In this work, we propose the existence of a previously unrecognized type of nodal loops, which appear when $|\bm w_\bot|$ becomes large enough such that the tilt term dominates the spectrum in Eq.~(\ref{Epm}). This happens when
$|\bm w_\bot|^2>\sqrt{v_1^2 w_1^2+v_2^2 w_2^2}$. In such case, the spectrum becomes completely tipped over along the $\bm w_\bot$ direction [Fig.~\ref{fig1}(c)], where the two crossing bands now have the \emph{same} sign for their slopes. (Note that for directions away from $\bm w_\bot$, the spectrum may still be of the usual type.) Parallel to the discussion in the context of nodal points~\cite{Soluyanov2015,Xu2015}, we term such type of loops as type-II, to distinguish them from the conventional (type-I) nodal loops.

We first point out that type-II nodal loops could share the same protection mechanisms as their type-I counterparts. The Berry phase is still well-defined, although there may not be a \emph{global} gap along the loop $\ell$ (local gap at each point on $\ell$ is sufficient for a well-defined Berry phase). Indeed, the winding of the pseudospin $\bm\sigma$ is not affected by the tilt, which only acts as a $\bm q$-dependent overall energy shift.

Consider the case when the loop lies in a mirror plane. Then the tilt vector $\bm w$ is constrained to be in this plane. Assuming that $\bm w$ is along the $q_1$ direction, the condition for a type-II (type-I) nodal loop becomes $|\bm w|>|v_1|$ ($|\bm w|<|v_1|$). One simplest model that describes a single nodal loop may be written as
\begin{equation}\label{Hring}
H=\frac{1}{2m}k_\rho^2+\frac{1}{2\eta}(k_\rho^2-k_0^2)\sigma_x+v_zk_z\sigma_y,
\end{equation}
where $k_\rho=\sqrt{k_x^2+k_y^2}$, $m$, $\eta$, and $k_0$ are model parameters. The model describes a nodal loop with radius $k_0$ in the $k_z=0$ plane. Evidently, it contains (\ref{Pmodel}) as a low-energy model via identifying $q_1$ to be along the in-plane radial direction $\hat{k}_\rho$, and $q_2$ to be along $\hat k_z$-direction, with the correspondences that $v_1=k_0/\eta$, and $\bm w=k_0/m\hat k_\rho$. Hence the loop is type-II (type-I) when $|\eta/m|>1$ ($<1$). One observes that in the type-I case, the loop is formed by the crossing between an electron-like band and a hole-like band. Whereas for $|\eta/m|>1$, the tilt term (first term in Eq.~(\ref{Hring})) dominates, making both bands electron-like or hole-like along the radial direction depending on the sign of $m$, and their intersection makes the type-II nodal loop.

\begin{figure}[t!]
\includegraphics[width=8.6cm]{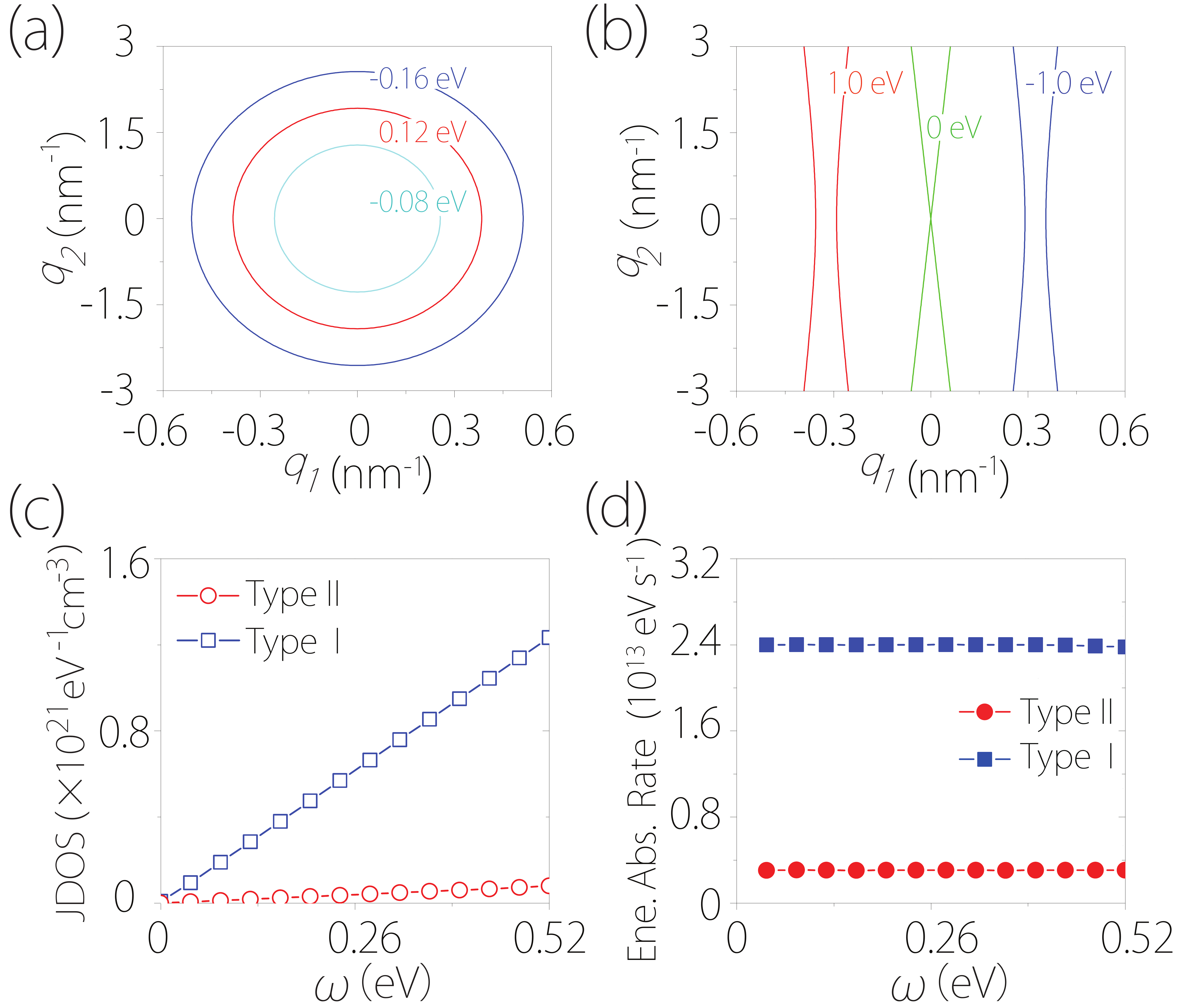}
\caption{Equi-energy contours in the $q_1$-$q_2$ plane for (a) type-I and (b) type-II loops. Comparison of (c) JDOS and (d) optical energy absorption rate for the two types of loops. Here we used model (\ref{Hring}) with parameters $k_0=1.7$ nm$^{-1}$, $\eta=0.4m_e$ ($m_e$ is the free electron mass), and $v_z=1\times 10^5$ m/s. We take $m=0.04m_e$ for type-II case and $m\rightarrow\infty$ for type-I case. Fermi level is set at the loop's energy for each case. In (d), the light $E$-field is polarized along $y$-direction with a peak value of 0.1 mV/nm.}
\label{fig2}
\end{figure}

The distinction between type-II and type-I loops can be observed from the geometry of their constant energy surfaces, which are shown in Fig.~\ref{fig2}(a,b) for the $q_1$-$q_2$ plane ($k_\rho$-$k_z$ plane) intersecting the ring. One observes that for type-I loop, the equi-energy contours are closed ellipses encircling the loop. When the Fermi energy is aligned to that of the loop (set as zero energy in the figure), the Fermi surface is simply given by the loop. In contrast, the equi-energy contours for type-II loop become hyperbolas~\cite{open}, and at the energy of the loop, the contour coincides with the two asymptotes.

The qualitative difference between the shapes of their constant energy surfaces will manifest in a variety of physical properties. For example, under a magnetic field, electrons orbit around constant energy surfaces, the different types of orbits will produce contrasting signals in magneto-oscillations~\cite{OBrien2016,Li2016a,Khim2016} (such as de Haas-van Alphen oscillations); and the transition from elliptic to hyperbolic type orbits, e.g., by varying the magnetic field direction for a type-II nodal loop, would typically be accompanied by a Landau level collapse phenomenon~\cite{Yu2016a}.

One notes from Fig.~\ref{fig2}(a,b) that the positive and negative energy contours have a much less overlap for type-II loop than for type-I. This is a natural consequence of both bands being electron-like (or hole-like), and could lead to marked difference in their optical response. Assuming the Fermi level is at the loop's energy, optical absorption involves transitions from negative energy states to positive energy states at the \emph{same} $k$-point. Compared to the type-I case, the positive and negative energy states for a type-II loop are largely separated in $k$-space, leading to much smaller absorption at low energies. This difference can be inferred from the joint density of states (JDOS): $\mathcal{D}(\omega)=\frac{1}{V}\sum_{\bm k}\delta(E_{c,\bm k}-E_{v,\bm k}-\omega)$, and the optical energy absorption rate: $\mathcal{R}(\omega)={2\pi}\omega\sum_{\bm k}|M_{cv}|^2 \delta(E_{c,\bm k}-E_{v,\bm k}-\omega)$. Here $M_{cv}$ is the optical transition matrix element, and note that $E_c$ ($E_v$) is for states above (below) the Fermi level. In Fig.~\ref{fig2}(c) and \ref{fig2}(d), we plot JDOS and $\mathcal{R}$ (for light with linear polarization in the $xy$-plane) calculated for the model in Eq.~(\ref{Hring}). One indeed observes that both quantities are much suppressed for the type-II case.

We also briefly remark that for carrier transport in the plane of the loop, the type-II loop may have a higher mobility than the type-I case. This is because that while they both share the enhancement due to $\pi$ Berry phase~\cite{Mullen2015}, the low-energy states near a point on the type-II loop are propagating roughly at the same direction, while the opposite-propagating states are located at the other end of the loop (c.f. Fig.~\ref{fig1}), thus momentum relaxation by scattering would be less efficient as compared with the type-I case~\cite{blackhole}.

\begin{figure}[t!]
\includegraphics[width=8.8cm]{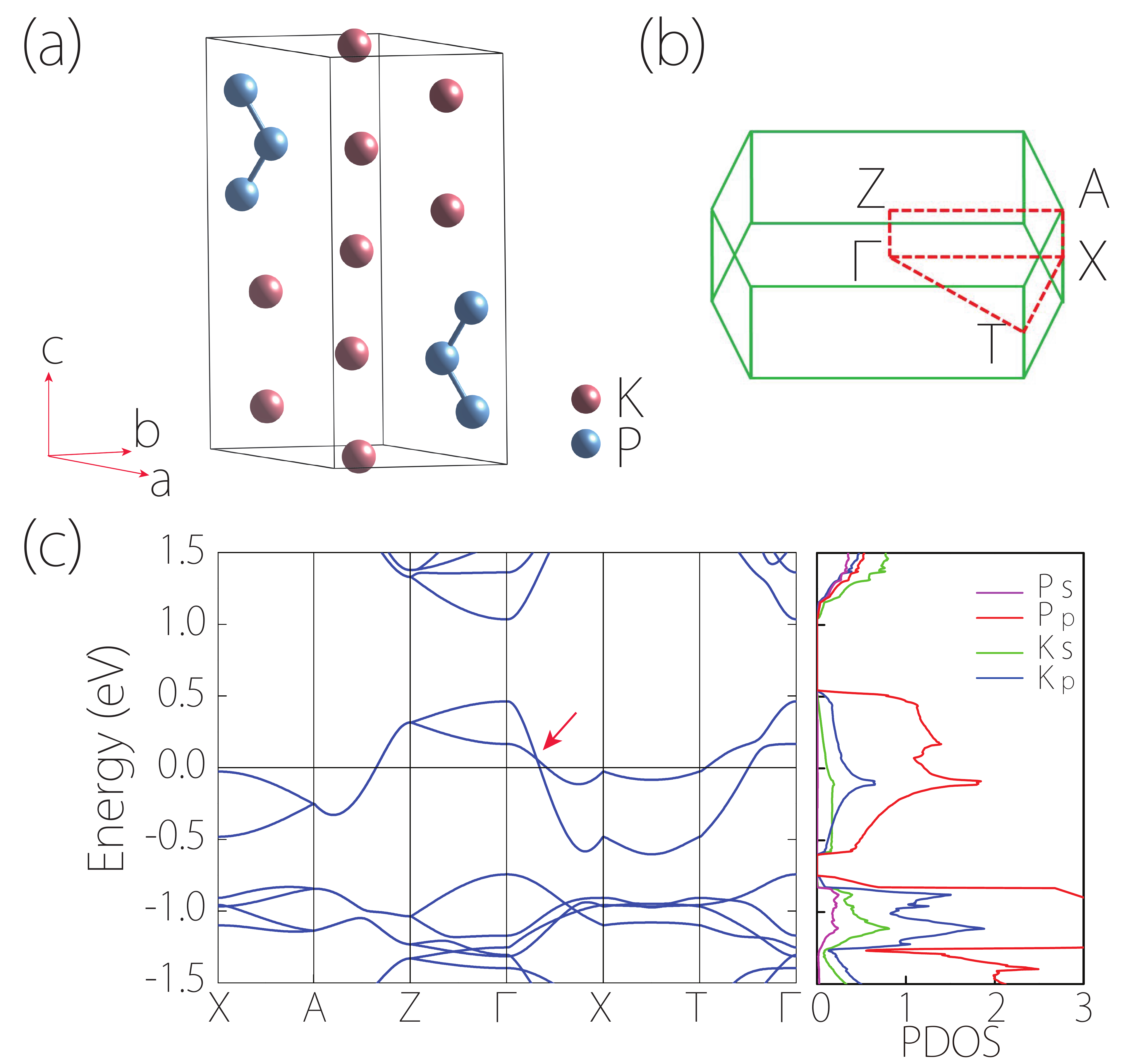}
\caption{(a) Crystal structure of K$_4$P$_3$. (b) Brillouin zone with high symmetry points labeled. (c) Electronic band structure of K$_4$P$_3$ and the projected density of states (PDOS). The red arrow indicates the crossing point on a type-II nodal loop.  }
\label{fig3}
\end{figure}

We now describe a concrete material realization for the type-II nodal loops---the crystalline compound K$_4$P$_3$. The single crystal K$_4$P$_3$ solid has been synthesized experimentally through reaction of red phosphorous with excess potassium~\cite{Schnering1989b}. The material is a stable paramagnetic metal at ambient condition, taking a W$_3$CoB$_3$-type orthorhombic structure with space group No. 63 (\emph{Cmcm})~\cite{Schnering1989b} [see Fig.~\ref{fig3}(a)]. The structure has angular P$_3$-chains, with each P atom lying in the center of a trigonal prism formed by six K$^+$ ions~\cite{Sangster2010,Poettgen2011}. The detailed structure information can be found in Ref.~\cite{Schnering1989b}, and experimental lattice parameters ($a=b=6.141$ \AA, $c=14.788$\AA)~\cite{Schnering1989b} are used in the calculation. For the following discussion, it is important to note the presence of two symmetries: the inversion symmetry $\mathcal{P}$ and the glide mirror symmetry $\mathcal{M}$ about $(110)$ plane [with $(x,y,z)\rightarrow (x,-y,z+1/2)$].

We performed first-principles calculations based on the density functional theory (DFT), as implemented in the Vienna Ab-initio Simulation Package~\cite{Kresse1994,Kresse1996}. The ionic potentials were modeled with the projector augmented wave method~\cite{PAW}, and the exchange-correlation functional was approximated in the generalized gradient approximation with the Perdew-Burke-Ernzerhof (PBE) realization~\cite{PBE}. The cutoff energy was chosen as 400 eV, and the Brillouin zone (BZ) was sampled with a $\Gamma$-centered $k$-mesh of size $12\times 12\times 6$. The energy and force convergence criteria were set to be $10^{-5}$ eV and $0.01$ eV/\AA, respectively. The band structures with and without SOC show very little difference, hence SOC is neglected in the following discussion. The surface states were investigated using the method with maximally localized Wannier functions~\cite{Marzari1997,Souza2001,Wu_tool}.

\begin{figure}[t!]
\includegraphics[width=9cm]{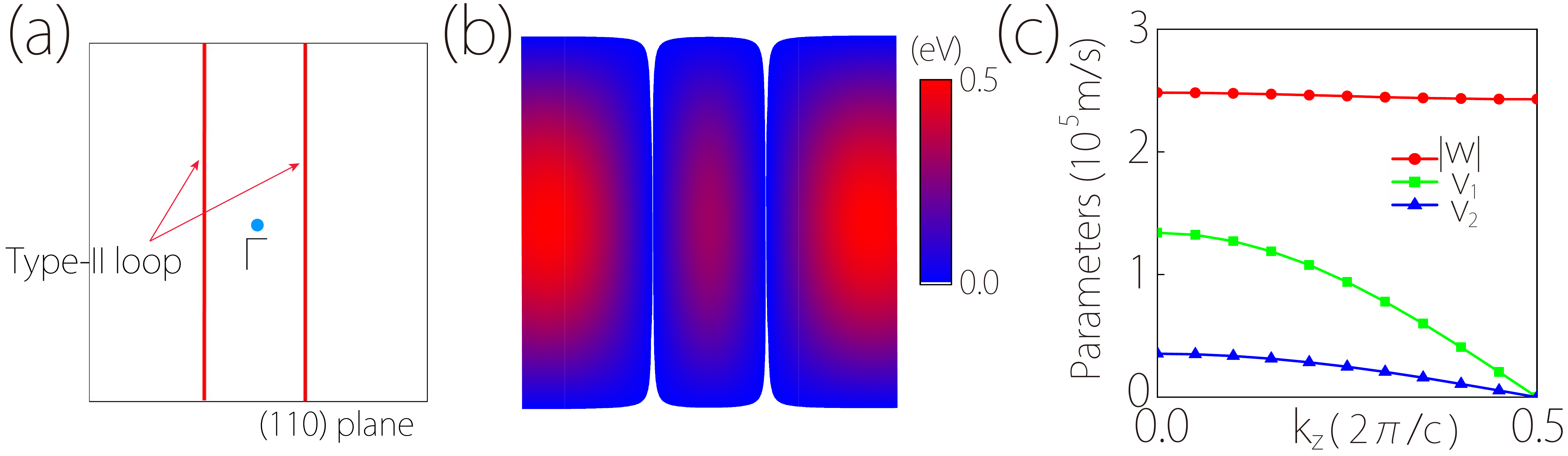}
\caption{(a) Schematic figure showing the location of the type-II loops in the (110) plane, and (b) shows the corresponding result from DFT. The color-map shows the local gap between the two crossing bands. (c) Parameters of effective model (\ref{Pmodel}) obtained by fitting the DFT band structure. }
\label{fig4}
\end{figure}

The band structure of K$_4$P$_3$ is shown in Fig.~\ref{fig3}(c). One observes that the system is metallic, and from the projected density of states (PDOS), the low-energy states are mainly from the $p$-orbitals of P atoms. There are two low-energy bands, which cross each other linearly along $\Gamma$-X, forming a crossing point as indicated in Fig.~\ref{fig3}(c). Since the system preserves both $\mathcal{P}$ and $\mathcal{T}$ symmetries, which dictates a vanishing Berry curvature field~\cite{Xiao2010}, this linear crossing point cannot be isolated. Indeed, a careful scan of the band structure reveals that the crossing between the two bands form a pair of nodal loops, as illustrated in Fig.~\ref{fig4}(a,b). The two loops are lying in the (110) plane, as constrained by the $\mathcal{M}$ symmetry, and they are quite straight. The energy variation along the loops is small ($<0.01$ eV), and the loops can be brought even closer to the Fermi level by doping or applying pressure~\cite{Supp}. Interestingly, each loop traverses the whole BZ, a feature that has important consequences to be discussed in a while.

Most important for our discussion is the observation that the dispersion around the loop is of type-II in the mirror plane along the $[\bar{1}10]$ direction. Model (\ref{Pmodel}) can be used to fit the DFT band structure.
Since each loop is almost a straight line along $k_z$, $q_1$ and $q_2$ can be taken as orthogonal components along $[\bar{1}10]$ and $[110]$ directions respectively. The tilt vector $\bm w$ is in the $q_1$ direction as required by $\mathcal{M}$, and its sign is opposite for the two loops. The fitted parameters are plotted in Fig.~\ref{fig4}(c). The value of $|w|$ slightly varies around $2.5\times 10^5$ m/s, while both $v_1$ and $v_2$ are maximum at $k_z=0$ and approach zero towards the Brillouin zone boundary.
Most importantly, one observes that $|w|>|v_1|$ for the whole loop, therefore, the loop is type-II.

The type-II loops here have two independent symmetry protections. One protection is from the  $\mathcal{P}$ and $\mathcal{T}$ symmetries, which ensures a quantized $\pi$ Berry phase for any close path encircling each loop. The other protection is from $\mathcal{M}$ since the two crossing bands have opposite $\mathcal{M}$ eigenvalues, as we have checked in DFT result. Consequently, the loop is stable against perturbations as long as one of the two protections is preserved.

\begin{figure}[t!]
\includegraphics[width=8.6cm]{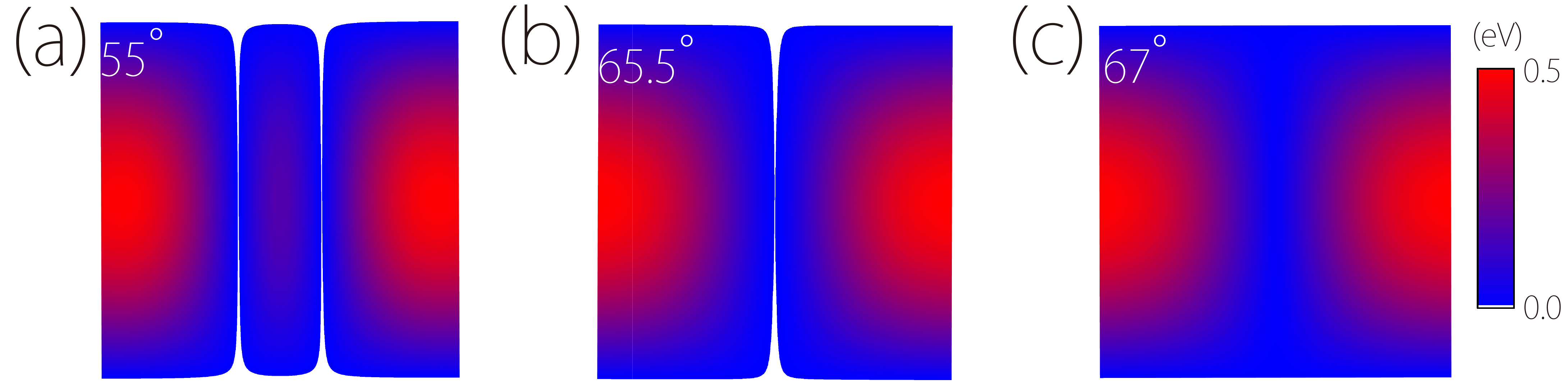}
\caption{Evolution of the loops when changing the angle $\gamma$ between $a$ and $b$ axis, for (a) $\gamma=55^\circ$,  (b) $\gamma=65.5^\circ$, and (c) $\gamma=67^\circ$. }
\label{fig5}
\end{figure}

Here each type-II loop is traversing the whole BZ. Such kind of loop is topologically distinct from those which are not penetrating the BZ, because the former cannot be continuously contracted to a point~\cite{Yang2016}, whereas the latter can. Mathematically, the BZ is topologically equivalent to a three-dimensional torus $\mathbb{T}^3$. Closed loops on $\mathbb{T}^3$ can be classified under its fundamental homotopy group $\pi_1(\mathbb{T}^3)=Z^3$, labeled by three integers, each indicating the number of times the loop winds around one of the three directions. In this sense, the nodal loops not traversing BZ belong to the trivial class with $Z^3=(0,0,0)$ (which includes a single point), whereas the loops here belong to the $(0,0,1)$ class. Hence the two kinds of loops cannot be continuously deformed into each other. This also means that with preserved symmetry, each of the two loops here cannot be annihilated by itself; they can only annihilate in pair. One such scenario is shown in Fig.~\ref{fig5}, where we vary the angle $\gamma$ between $a$ and $b$ axis, which preserves the crystal symmetry. With increasing $\gamma$, the two loops are moving towards the BZ center and finally annihilate with each other.

\begin{figure}[t!]
\includegraphics[width=8.6cm]{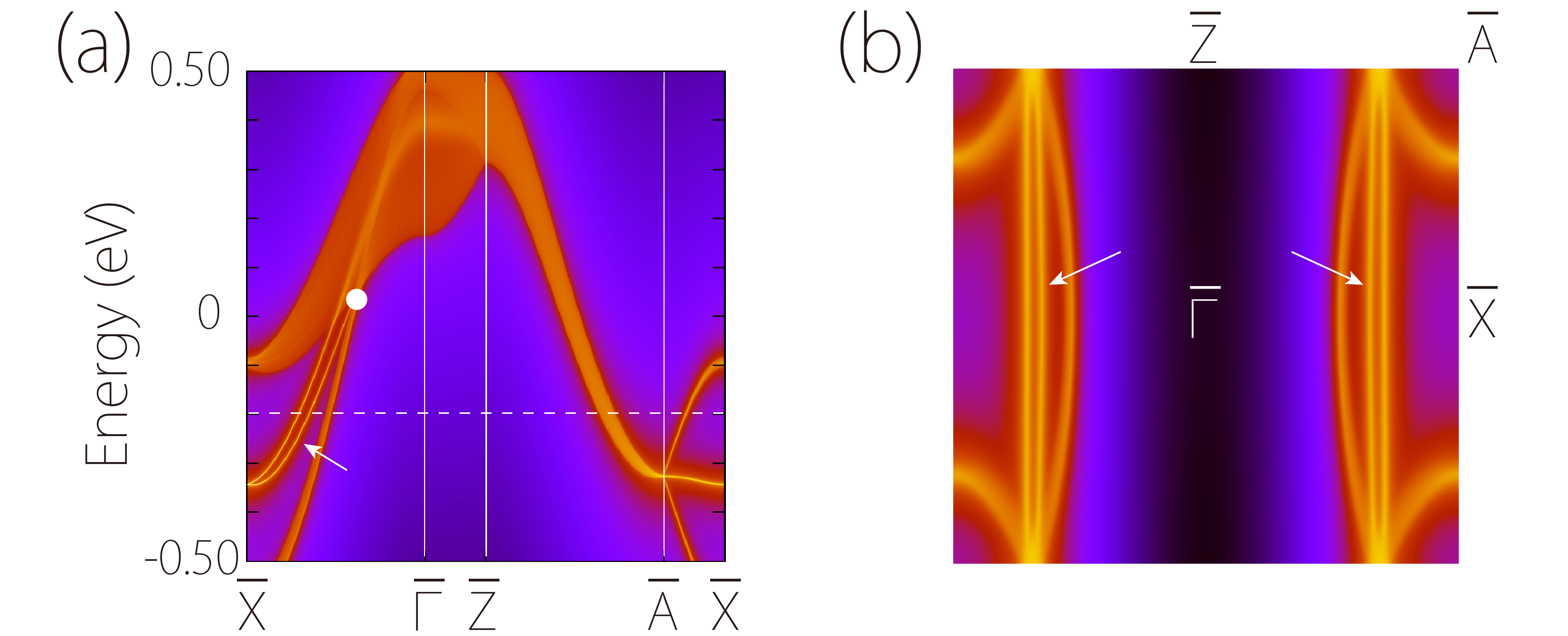}
\caption{(a) Projected spectrum on (110) surface, and (b) the corresponding constant energy slice at -0.2 eV. The white dot in (a) marks the projected bulk band crossing point. The arrows indicate the drumhead-like surface states.  }
\label{fig6}
\end{figure}

Nodal loops usually possess drumhead-like surface states~\cite{Weng2015c}. In Fig.~\ref{fig6}, we show the spectrum of the (110) surface of K$_4$P$_3$. Indeed, one observes the drumhead surface band emanated from the bulk nodal points. The surface band connects the two loops through the surface BZ boundary. We verify that each bulk line along [110] direction and in the surface band region carries a quantized $\pi$ Berry phase, hence contributing a state at the surface terminated by vacuum~\cite{Supp}. One also notes an additional surface band with slightly higher energy which we find is originated from the surface dangling bonds.

Before closing, we mention that the type-II nodal loops in K$_4$P$_3$ and the associated surface states can be directly probed via angle-resolved photoemission spectroscopy. In addition, the type-II nature of the loops may also manifest in the contrast between DOS and JDOS, and in the magnetic response of K$_4$P$_3$, as discussed in \cite{Supp}.
We also point out that besides type-I and type-II loops, there could also be a hybrid type for which the tilt vector dominates only over part of the loop. In terms of physical properties, the hybrid type should be intermediate between type-I and type-II. We find real materials that possess such hybrid loops, e.g., in the ScCd-type transition-metal intermetallic materials~\cite{hybridloop}. Hybrid nodal lines connecting nexus points have also been predicted in Bernal stacked graphite~\cite{Heikkilae2015,Hyart2016,KZhang}. Finally, for K$_4$P$_3$, the direction of tilt vector is pinned onto the glide mirror plane, but for systems with reduced symmetry, the tilt vector may wind around when going along the loop. How such variation would affect physical properties could be an interesting topic to investigate in future works.



\begin{thebibliography}{58}
\bibitem [{ope()}]{open}%
  \BibitemOpen
  \href@noop {} {}\bibinfo {note} {The open contour in the effective model can
  become closed when the model is regularized on a lattice.}\BibitemShut
  {Stop}%
\end{thebibliography}
%

\end{document}